\documentclass[twocolumn]{article}
\usepackage[utf8]{inputenc}
\pdfoutput=1

\usepackage[left=1.5cm, right=1.5cm, top=1.785cm, bottom=2.0cm]{geometry}
\usepackage{amsmath}
\usepackage{siunitx}
\DeclareSIUnit{\molar}{M}

\usepackage{graphicx} 
\graphicspath{{figures/},{finalfig/}}

\usepackage{lmodern}
\usepackage[english]{babel}
\usepackage{array}
\usepackage{array,multirow}
\usepackage[T1]{fontenc}
\usepackage[usenames,dvipsnames]{xcolor}
 
\usepackage{soul}
\usepackage[format=plain,justification=justified,singlelinecheck=false,font={footnotesize,sf},labelfont=bf,labelsep=space]{caption}
\usepackage{float}
\usepackage{csquotes}
\usepackage{cleveref}
\crefname{equation}{eqn}{eqn}
\crefname{figure}{Fig.}{Figs.}
\crefname{table}{Table}{Tables}

\newcommand\blfootnote[1]{%
  \begingroup
  \renewcommand\thefootnote{}\footnote{#1}%
  \addtocounter{footnote}{-1}%
  \endgroup
}
\usepackage{multirow,bigdelim}
\usepackage{booktabs}
\newcommand{\gcos}{g}

\newcommand{\G}{\Gamma}
\newcommand{\x}{_{\mbox{\tiny xlink}}}
\newcommand{\mer}{_{\mbox{\tiny mer}}}
\newcommand{\tot}{_{\mbox{\tiny tot}}}
\newcommand{\cs}{_{\mbox{\tiny solute}}}

\usepackage{multirow}
\usepackage{gensymb}

\usepackage{cleveref}
\crefname{equation}{eqn}{eqn}
\crefname{figure}{Fig.}{Figs.}
\crefname{table}{Table}{Tables}

\renewcommand{\footnoterule}{%
  \kern -3pt
  \hrule width \linewidth height 0.5pt
  \kern 2pt
}

\usepackage[backend=biber,natbib=true,sorting=none]{biblatex}
\bibliography{bib/bib2.bib} 
\renewcommand{\cite}{\supercite }

\definecolor{mycolorNP0}{rgb}{1.0000,0.60000,0.10000}%%
\definecolor{mycolorwat}{rgb}{0.0500,0.50000,1.0000}%%
\definecolor{mycolorMeO}{rgb}{1.000,0.30000,0.5000}%%

\definecolor{mycolor1}{rgb}{0.00000,1.00000,1.00000}%
\definecolor{mycolor2}{rgb}{0.70000,0.00000,1.00000}%

\definecolor{mycolorg1}{rgb}{0.00000,0.44700,0.74100}%
\definecolor{mycolorg2}{rgb}{0.85000,0.32500,0.09800}%
\definecolor{mycolorg3}{rgb}{0.75,1,0.65}%

\definecolor{cream}{RGB}{222,217,201}

\date{\small\textit{$^{a}$~Research Group for Simulations of Energy Materials,\\ Helmholtz-Zentrum Berlin f\"ur Materialien und Energie,\\Hahn-Meitner-Platz 1, D-14109  Berlin,  Germany. }\\
\textit{$^{b}$~Institut f\"ur Physik, Humboldt-Universit\"at zu Berlin,\\ Newtonstr. 15, D-12489 Berlin, Germany. }\\
\textit{$^{c}$~Applied Theoretical Physics  - Computational Physics, Physikalisches Institut,\\Albert-Ludwigs-Universit\"at Freiburg, Hermann-Herder Str. 3, D-79104 Freiburg, Germany. }\\
\textit{$^{d}$~Jo\v{z}ef Stefan Institute, Jamova 39, SI-1000, Ljubljana, Slovenia. }
}

\title{\textbf{Cross-linker effect on solute adsorption in swollen thermoresponsive polymer networks}}
\author{Sebastian Milster,\textit{$^{a,b,c}$} Richard Chudoba,\textit{$^{a,b,c}$}\\ Matej Kandu\v{c}\textit{$^{a,d}$} and Joachim Dzubiella$^{\ast}$\textit{$^{a,c}$}}

\begin{document}
\thispagestyle{empty}

\twocolumn[
  \begin{@twocolumnfalse}
  \maketitle
  \begin{center}
   \begin{tabular}{p{14cm}}

The selective solute partitioning within a polymeric network is of key importance to applications in which controlled release or uptake of solutes in a responsive hydrogel is required. In this work we investigate the impact of cross-links on solute adsorption in a swollen polymer network by means of all-atom, explicit-water molecular dynamics simulations. We focus on a representative network subunit consisting of poly($N$-isopropylacrylamide) (PNIPAM) and $N$,$N'$-methylenebisacrylamide (BIS/MBA) cross-linker types. Our studied system consists of one BIS-linker with four atactic PNIPAM chains attached in a tetrahedral geometry. The adsorption of several representative solutes of different polarity in the low concentration  limit at the linker region is examined. We subdivide the solute adsorption regions and distinguish between contributions stemming from polymer chains and cross-link parts. In comparison to a single polymer chain, we observe that the adsorption of the solutes to the cross-link region can 
significantly differ, with details depending on the specific compounds' size and polarity.  In particular, for solutes that have already a relatively large affinity to PNIPAM chains the dense cross-link region (where many-body attractions are at play) amplifies the local adsorption by an order of magnitude. We also find that the cross-link region can serve as a seed for the aggregation of mutually attractive solutes at higher solute concentrations. Utilizing the microscopic adsorption coefficients in a mean-field model of an idealized macroscopic polymer network, we extrapolate these results to the global solute partitioning in a swollen hydrogel and predict that these adsorption features may lead to non-monotonic partition ratios as a function of the cross-link density.
  \end{tabular}
  \end{center}

 \end{@twocolumnfalse} \vspace{0.6cm}
  ]
\blfootnote{\footnotesize E-mail: \texttt{joachim.dzubiella@physik.uni-freiburg.de}}

 \section{Introduction}
 
Responsive polymers have increasingly gained attention in many research fields due to their ability to reversibly adapt to external stimuli such as temperature, osmotic pressure, or pH. Various types and shapes at different length scales have been designed, providing various possibilities for applications,\cite{Stuart2010,fernandez2011microgel} such as solute uptake, transport\cite{bae1991off} and release,\cite{Motornov2010,schild1992poly}  sensors,\cite{hendrickson2009bioresponsive} intelligent coatings, switchable catalysis,\cite{roa2017catalyzed,carregal2010catalysis} etc. To structurally stabilize and fine-tune properties and function, polymer architectures are often equipped with chemical cross-linkers covalently interconnecting the chains,\cite{hennink2012novel} which are then referred to as hydrogels. Typical responsive hydrogels in most studies barely exceed a molar cross-linker density of twenty percent, since greater values increase the rigidity of the gel and reduce the swelling properties due to the denser polymer network structure.\cite{zhu2012preparation,carregal2010catalysis}
 
\begin{figure}
\begin{center}

\includegraphics[width=\linewidth]{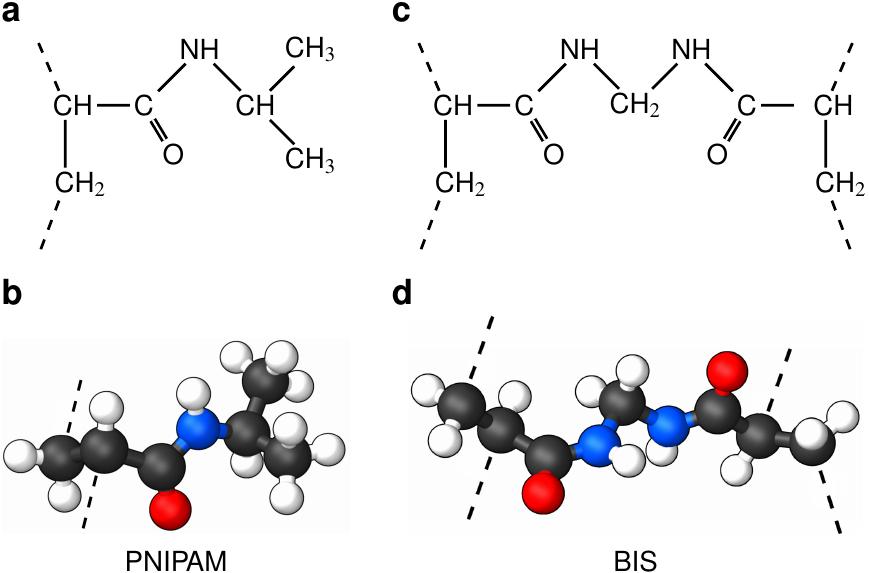}
 \end{center}
 \caption{Chemical structures (top) and corresponding ball-and-stick representations (bottom) of poly($N$-isopropylacrylamide) [PNIPAM] (panels {\bf\sffamily a}, {\bf\sffamily b}),  and the cross-linker $N$,$N'$-methylenebisacrylamide [BIS] (panels {\bf\sffamily c}, {\bf\sffamily d}). Dashed lines represent possible bonds to neighboring PNIPAM monomers or BIS. Associated carbon atoms are referred to as the polymer backbone. PNIPAM's amide group and isopropyl group form side chains. The amide groups and BIS' central methylene bridge are hydrophilic, potentially forming hydrogen bonds{\cite{luzar1996hydrogen}} with the surrounding. The backbone and isopropyl group have a hydrophobic character.\label{fig:chemstruct}}
 \end{figure}

In the zoo of constituents, thermoresponsive hydrogels based on poly($N$-isopropylacrylamide) (PNIPAM) are among the most intensively investigated systems, since their volume phase transition at about room temperature as well as a high water content promise good biocompatibility\cite{Gehrke1997,schild1992poly} and make them convenient to handle. Pure PNIPAM was found to have the lower critical solution temperature (LCST) at roughly $\SI{304}{\kelvin}$ as reported by Heskins and Guilett in 1968.\cite{Heskins1968} A frequently utilized cross-linker for PNIPAM gels, used in radical polymerization, is $N$,$N'$-methylenebisacrylamide, often abbreviated as BIS or MBA. BIS has chemical similarity to PNIPAM (compare \cref{fig:chemstruct}), is non-degradable, has a very high reactivity, and retains PNIPAM's LCST.\cite{Sanson,still2013synthesis,Nagaoka1993,Hirokawa1984,senff2000influence} 

Besides these morphological properties, the degree of cross-linking influences solute uptake and partitioning. The partition ratio is the ratio of the solute concentrations inside and outside the gel and is therefore a crucial parameter controlling device functionality especially for drug delivery or catalytic systems. For the latter, for instance, metal nanoparticles inside hydrogels catalyze reactions and the effective reaction rates depend crucially on the concentration of the reactants in the permeable polymer matrix.\cite{herves2012catalysis,carregal2010catalysis, roa2017catalyzed,Lu2011}  The partition ratio may be affected by generic as well as specific cross-linker effects. The cross-linker density first of all simply changes the packing fraction and with that the overall steric exclusion by the polymer mesh.\cite{Gehrke1997} Furthermore, it has become clear that more complex, e.g., local attractive and/or electrostatic interactions can lead to complex and even cooperative effects in 
the partitioning.\cite{kim2017cosolute,Irene, perez2018maximizing, kosovan1} In particular,  a `vertex trapping' effect due to many-body attractions in the dense cross-link region has been reported in generic coarse-grained simulations of polymer networks.\cite{Zhang2015,hansing2016nanoparticle,hansing2018hydrodynamic,hansing2018particle,hansing2018particle2} More specific chemical effects should also play a role, as indicated by all-atom molecular dynamics (MD) computer simulations of bare PNIPAM chains\cite{Alaghemandi2012,deshmukh2012role,janOPLSQM2}, peptide-like chains\cite{schwierz2012relationship,schwierz2016mechanism,kienle2012effect}, these in combination with solutes with various polarity \cite{du2010effects,horinek,rodriguiez-ropero,matej,matej2, nayar} as well as by simulations revealing the influence of cross-links to polymer networks solvation and structural properties. \cite{tonsing2001molecular,deshmukh2009,deshmukh2011,kosovan2} 

The aim of this work is to investigate the effects of cross-linking on solute adsorption in swollen hydrogels made up of PNIPAM and BIS (below the {volume phase transition temperature (VPTT)}) by utilizing all-atom, explicit water MD simulations of a minimal polymer network setup. In order to do this, we consider one BIS cross-linker with four atactic PNIPAM chains restrained in a tetrahedral geometry. In our analysis, we subdivide the solute adsorption regions and systematically distinguish between contributions stemming from polymer chains and cross-linker parts. We probe solutes of various polarity, representing typical chemical compounds,  in the highly diluted regime. We finally demonstrate in a simple model, how these contributions affect the global solute partitioning in large hydrogels as accessible by experiments.

 \section{Methods}

\subsection{Hydrogel building blocks: PNIPAM and BIS}

Constructing covalently cross-linked polymeric networks for our computer simulations requires two types of building blocks: chain monomers and the cross-linker (see \cref{fig:chemstruct}). The former provides two binding sites and builds up the chains. The latter has four binding sites and thus interconnects four chains. We have chosen poly($N$-isopropylacrylamide) (PNIPAM) and $N$,$N'$-methylenebisacrylamide (BIS) for the chains and the cross-linker unit respectively (see \cref{fig:chemstruct}).

We employ the OPLS-QM2 force field recently developed by Palivec et al.\cite{janOPLSQM2} for the PNIPAM monomers. Compared to the standard OPLS-AA \cite{jorgensen_opls_1988} parameters, this force field features a reparametrization of the partial charges retrieved from ab-initio calculations and further manual fine-tuning to reproduce the experimental LCST of PNIPAM. Due to the chemical similarity of BIS and PNIPAM, we adopt the very same partial charges for the cross-linker. These were confirmed by our own quantum mechanical calculations using the Gaussian 09 software. \cite{g09} More details on the force field parameters are provided in \Cref{sec:appA}. 

\subsection{Setup}

 \begin{figure*}
 \centering
\includegraphics{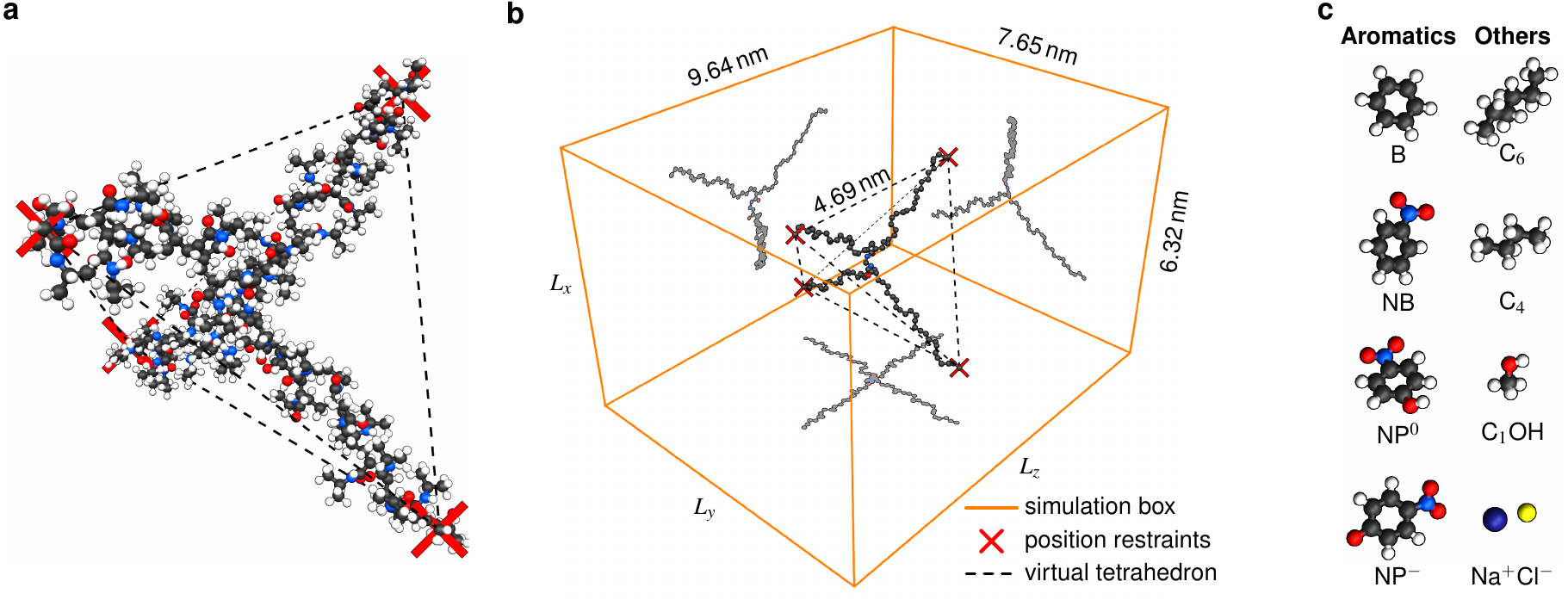}

 \caption{Simulation snapshots of the studied polymeric molecule, consisting of one BIS-linker and four PNIPAM chains, at $T=\SI{290}{\kelvin}$ and relative chain extension $\lambda=0.83$ in ({\bf\sffamily a}) all-atom ball-and-stick representation and ({\bf\sffamily b}) showing only backbone and heavy BIS atoms inside the simulations box ({\color{orange}\bf -----}) and in addition as 2D projections. Terminal backbone carbons are position-restrained (marked as {\color{red}$\bf\times$}) in the corners of a virtual tetrahedron ({\color{black}\bf - - -}), which has an edge length of \SI{4.69}{\nano\meter} and a vertex-to-centroid distance of $L=\SI{2.87}{\nano\meter}$. For the sake of clarity, water molecules are not shown. ({\bf\sffamily c}) Probe solutes.  Atoms are color-coded by element, i.e., black carbon, red oxygen, blue nitrogen, and white hydrogen atoms.   See \cref{tab:sim_time} for the full simulation specifications.  \label{fig:setup}}
\end{figure*}
The actual polymerization procedure in experiments is subjected to randomness leading to different possible network topologies. Looking at the very local structure inside a hydrogel, the following sources for inhomogeneities are possible: dangling chains, entangled chains, and loops.\cite{ikkai2005inhomogeneity} This work, however, focuses on generic subunits of a defect-free network architecture of a swollen hydrogel,\cite{perez2018maximizing,schneider2002swelling} namely the cross-linker and its four adjacent chains. Our setup consists of one BIS-linker connected with four PNIPAM chains (\cref{fig:setup}a), each composed of 12 monomers, i.e., 48 monomers in total. The terminal backbone carbon atoms of each chain are position-restrained in order to sustain a tetrahedral structure. It further facilitates a clear analysis and the results can be to some extent generalized, which will be discussed later in this work.

The corners of the tetrahedron, generated by the position restraints, are $L=\SI{2.87}{\nano\meter}$ distant to its centroid and the edge length accounts to $\SI{4.69}{nm}$ (\cref{fig:setup}b). We have chosen this size to ensure a relative chain stretch $\lambda$ between $0.75$ and $0.85$, which is expected in swollen hydrogels.\cite{matej} The relative chain stretch $\lambda$ is defined as the ratio of the mean end-to-end distance and the contour length
 \begin{align}
  \lambda=\frac{\left< \ell_{\mbox{\tiny ee}}\right>}{L_{\mbox{\tiny c}}}.
 \end{align}
The mean end-to-end distance of the chains is $\left< \ell_{\mbox{\tiny ee}}\right>=\SI{2.65}{\nano\meter}$, where the brackets $\langle..\rangle$ denote the ensemble average. The distance $\ell_{\mbox{\tiny ee}}$ was measured from the cross-linker contact backbone atom to the chain's terminal backbone atoms during the $NpT$-simulation. The contour length per monomer $\Delta L_{\mbox{\tiny c}}$ in an atactic PNIPAM chain is approximately $\SI{0.265}{\nano\meter}$, and when multiplied by 12, one obtains the contour length of a single chain.  Eventually, the average relative stretch in all simulations is $\lambda={0.83}$.

Generating such a setup starts by placing the BIS-linker in the center of the box, which is the centroid of the virtual tetrahedron. At each backbone binding site, indicated by dashed lines in \cref{fig:chemstruct}, PNIPAM monomers with random tacticity are attached in a head-to-tail manner with the backbone axis pointing towards the desired coordinates of the position restraint. PNIPAM's backbone bonds are squeezed to match the size of the tetrahedron and were allowed to relax during the first steps (energy minimization and equilibration) of the simulation.

The edge lengths of the rectangular simulation box are chosen large enough ($\SI{6.32}{\nano\meter}\times\SI{7.65}{\nano\meter}\times\SI{9.64}{\nano\meter}$ on average) in order to ensure that the position-restrained backbone terminals of different periodic images are separated by at least \SI{3}{\nano\meter} in $x$- and $y$-, and \SI{5}{\nano\meter} in $z$-direction (\cref{fig:setup}b). Thus, we avoid interactions between chain ends across the box boundaries and can locate a (water/solute) bulk phase in peripheral box regions.

The solutes are subsequently inserted at random positions in the simulation box, which is finally filled with more than 15000 SPC/E \cite{spce} water molecules. Details are listed in \cref{tab:sim_time}.

\subsection{Probe molecules}
We analyze the adsorption properties of several types of molecules covering different sizes and polarity. We focus on some aromatic compounds due to their application as model reactants in catalytic  experiments \cite{herves2012catalysis,wu2012thermosensitive} and since aromatic rings are found in numerous drugs,{\cite{molina2012study,hoare2008impact,ritchie2009impact}} e.g., common painkillers. Precisely, these compounds are benzene (B), nitrobenzene (NB), (uncharged) 4-nitrophenol (NP$^0$), and (charged) deprotonated 4-nitrophenolate (NP$^-$).

We further probe two alkanes, namely hexane (C$_6$) and butane (C$_4$), sodium chloride (Na$^+$/Cl$^-$), and methanol (C$_1$OH). All compounds are visualized in \cref{fig:setup}c. If not stated otherwise, we insert one probe molecule into the system to analyze the infinite dilution limit. To estimate finite concentration effects of aromatic compounds we perform simulations with 20 solutes. List of solutes and the simulation setups are listed in the summary \cref{tab:sim_time}. The standard OPLS-AA\cite{jorgensen_opls_1988} force field was utilized except for the charged nitrophenolate NP$^-$, for which the excess charge was distributed among the molecule due to the mesomeric effect,\cite{matej} leading to higher polarity of the nitro group.

\subsection{Simulation details}
We employed all-atom, explicit-water molecular dynamics (MD) simulations to study the polymer--solute interactions. The simulations were performed using the Gromacs 5.1 software \cite{Hess2008,VanDerSpoel2005,Abraham2015} utilizing force fields mentioned above.

All {covalent bonds of hydrogens} were constrained with the LINCS \cite{Hess1997} algorithm. The cut-off distance for Lennard-Jones and short-range electrostatic interactions was set to \SI{1.0}{\nano\meter} while long range electrostatics was accounted for by the Particle Mesh Ewald (PME) method with cubic interpolation and a grid spacing of \SI{0.12}{\nano\meter}. \cite{Essmann1995}

	Periodic boundary conditions in all three directions were used and the simulations were carried out under constant temperature and pressure, which were controlled by the velocity-rescale thermostat (at $T=\SI{290}{\kelvin}, \tau_T = \SI{0.1}{\pico\second}$) and the Berendsen barostat (at $p=\SI{1}{\bar}$, $\tau_p = \SI{1}{\pico\second}$), respectively. \cite{Bussi2007,Berendsen1984}

	After the initial energy minimization (steepest descent), the system was equilibrated in the $NVT$ ensemble for \SI{2}{\nano\second} and in the $NpT$ ensemble for another \SI{10}{\nano\second}. The integration step of the leap-frog integrator was set to \SI{2}{\femto\second} and data were collected every 10 ps. The total simulation time $t_{\mbox{\tiny sim}}$ per solute is summarized in \cref{tab:sim_time}.

\begin{table*}
\footnotesize
\centering
\caption{Simulation specifications and resulting solute adsorption coefficients to the polymer setup with a relative chain extension $\lambda\approx0.83$ as depicted in \cref{fig:setup}. Each $NpT$-simulation was carried out at $T=\SI{290}{\kelvin}$, $p=\SI{1}{\bar}$, with more than 15000 water molecules, one solute (30 molecules and three pairs in the case of C$_1$OH and Na$^+$/Cl$^-$, respectively) and analyzed within a simulation time of $t_{\mbox{\tiny sim}}$. Nitro-aromatics were further tested with $N\cs=20$ molecules and we  added Na$^+$ counterions for charged NP$^-$. {Note the differences in the simulation time, e.g., the simulations with $N_{\mbox{\tiny solute}}=1$ were carried out approximately ten times as long as with 20 individuals in order to reach sufficient sampling quality.} Adsorption coefficients $\G^* = \G/\rho_0$  are split into contributions as described by \cref{eq:gamma_tot,eq:gamma_mer} and are normalized by solute bulk concentration $\rho_0$ {(Details are provided in the main text). For the NP$^-$/Na$^+$ pairs, only results for NP$^-$ are presented. The ions Na$^+$ and Cl$^-$ were simulated together but analyzed separately. Both types yield very similar results.} The most relevant results, the adsorption coefficients  $\G^*\x$ and $\G^*\mer$ to the cross-linkers and monomers, respectively, are visualized in \cref{fig:master_plot} \label{tab:sim_time}  } 
 \begin{tabular}{
l
l
S
S[separate-uncertainty,table-figures-uncertainty=1,table-figures-integer = 1,table-figures-decimal = 0]  
S[separate-uncertainty,table-figures-uncertainty=3,table-figures-integer = 3,table-figures-decimal = 1]
S[separate-uncertainty,table-figures-uncertainty=1,table-figures-integer = 1,table-figures-decimal = 0] 
S[separate-uncertainty,table-figures-uncertainty=1,table-figures-integer = 1, table-figures-decimal = 0]
S[separate-uncertainty,table-figures-uncertainty=1,table-figures-integer = 0, table-figures-decimal = 2]
S[separate-uncertainty,table-figures-uncertainty=3,table-figures-integer = 3, table-figures-decimal = 0]}

 \multicolumn{4}{c}{\bf\sffamily Simulation Specifications } & \multicolumn{5}{c}{\bf\sffamily Results} \\ \cmidrule(r){1-4}\cmidrule(l){5-9}
\multicolumn{2}{l}{\bf\sffamily Aromatics}&{ $N\cs$}&  \multicolumn{1}{c}{$t_{\mbox{\tiny sim}}$ [$\mu$s]}& {$\rho_0$ [mM]} & \multicolumn{1}{c}{ $\G^*\tot$ [nm$^3$]}  &{ $ \G^*\x$  [nm$^3$] } &{ $\G^*\mer$  [nm$^3$]} & \multicolumn{1}{l}{ $\G^*_{\mbox{\tiny end}}$  [nm$^3$]} \\
 \hline %% tot x mer end
 B\rule{0pt}{2ex}  & benzene & 1 &  12.6& 3.2\pm 0.1  &54 \pm 3 & -2\pm2 & 1.0\pm0.1 & 7\pm3\\
 NB & nitrobenzene & 1  &  11.9 & 2.8\pm 0.1& 124 \pm 6 & 5\pm2 & 2.3\pm0.2 & 10\pm6\\
 %NB$^\P$ & nitrobenzene & 1 & 7282 & 8.2  & {-} & {-} & 2.2 \pm 0.1 & {-} \\
 NB & nitrobenzene & 20  &  1.1 & 28.0\pm 2& 720\pm220 & 70\pm20 & 16.0\pm2 & -140\pm220\\
  NP$^0$ & 4-nitrophenol &1   & 12.3 & 2.8\pm 0.1& 121\pm7 & 21\pm3 & 2.1\pm0.2 & -2\pm7\\
NP$^0$ &  4-nitrophenol & 20   & 1.2 & 42.1\pm 2&280\pm60 & 100\pm20 & 4.6\pm0.5 & -40\pm60\\
 NP$^-$ & 4-nitrophenolate & 1 & 8.4 & 2.3\pm 0.1&270\pm20 & 43\pm8 & 4.5\pm0.4 & 10\pm20\\
  NP$^-$/Na$^+$  & 4-nitrophenolate + sodium & {20 pairs }  & 1.1 & 63\pm 1.0& 67\pm3 & 14\pm2 & 1.3\pm0.1 & -11\pm3\\
  
  \multicolumn{8}{l}{}\\
  \multicolumn{8}{l}{\bf\sffamily Others} \\
 \hline
 C$_6$ \rule{0pt}{2ex}  & hexane & 1  & 11.2 & 3.3\pm 0.1& 33\pm5 & -2\pm4 & 0.8\pm0.1 & -1\pm5\\
 C$_4$& butane & 1  & 7.4 & 3.4\pm 0.1&  25\pm2 & -1\pm2 & 0.44\pm0.05 & 5\pm2\\
 C$_1$OH & methanol & 30 & 3.5 & 107.3\pm 1.0&  -2\pm1 & 0\pm4 & -0.05\pm0.05 & 1\pm1\\
 Na$^+$/Cl$^-$ & sodium chloride  &{3 pairs \ }  & 8.4 & 11.5\pm 0.1& -33\pm2 & 5\pm1 & -0.65 \pm 0.02 & -6\pm2\\
 %H$_2$O & water &&&&&&&\\
 \end{tabular}

\end{table*}
\section{Analysis and discussion}

From the trajectories we calculate the center of mass (COM) positions of the cross-linker and the structure of the chain monomers, the solute(s) and water molecules around it. The distances $r$ between the COM of the cross-linker and the COM of molecules is used to obtain the normalized radial density distributions $\gcos(r) = \rho(r)/\rho_0$ for BIS--water and BIS--solute. The bulk phase concentration $\rho_0$  (see \cref{fig:wat_distro} and \cref{tab:sim_time}) is obtained from simulations {by calculating the number density along the $z$-axis and averaged in the region \SI{1.5}{\nano\meter} distant to the restrained polymer atoms.}

{For the 20 NP$^-$/Na$^+$ pairs, we only analyzed the nitrophenolate trajectories. In the case of the Na$^+$/Cl$^-$ simulation, each ion type was analyzed individually. The results for sodium and chloride ions are very similar and yield the same results within the range of our precision and thus are presented for either type in \cref{tab:sim_time}. }

 {Further, the (radial)} PNIPAM monomer number distribution $\rho\mer(r)$ is retrieved, which helps us to distinguish between different polymer adsorption domains (\cref{fig:wat_distro}). The solutes' distributions are the basis for calculating the solute--polymer adsorption in our setup as detailed below.  We demonstrate how the splitting of the adsorption into the chain and the cross-linker contributions is achieved and how this can be used to estimate partition ratios of an entire hydrogel.

\subsection{Polymer distribution}
\begin{figure}[b!]

\includegraphics{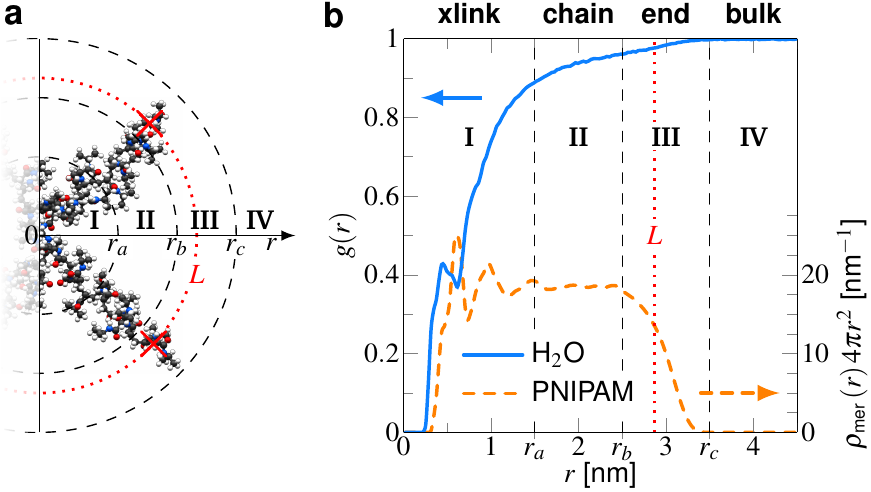}

\caption{The different adsorption domains {\bf I}, {\bf II}, and {\bf III} resolved in the radial distance $r$ to the COM of the cross-linker, illustrated in ({\bf \sffamily a}) a simulation snapshot of BIS plus two PNIPAM branches and ({\bf \sffamily b}) by radial water density (left hand axis) and PNIPAM monomer number (right hand axis) profiles. The first interval {\bf I} is the `cross-link region', where BIS as well as collective PNIPAM effects mix. The PNIPAM profile shows short-range oscillations, leading to the non-monotonicity in the water profile. The range of influence of the cross-linker is assumed to vanish around $r_a=\SI{1.5}{\nano\meter}$. Region {\bf II}, ranging from $r_a$ to $r_b$, is dominated by the linear PNIPAM chain. Here, the monomer number, $\rho\mer(r)4\pi r^2$, is roughly constant. In interval {\bf III}, from $r_b=\SI{2.5}{\nano\meter}$ to $r_c=\SI{3.5}{\nano\meter}$, the polymer chains terminate. The average distance between the position-restrained backbone terminals and the cross-linker, $L=\SI{2.87}{\nano\meter}$, is shown by the vertical dotted (red) line. For large distances, i.e., $r> r_c$, we assume negligible influence from the polymer and consider region {\bf IV} as `bulk'.
% The chain interval was chosen where PNIPAM monomer density is rather constant 
\label{fig:wat_distro}}
\end{figure}

The position restraints restrict the movement of the PNIPAM monomers and the cross-linker. The COM of the cross-linker in this setup fluctuates around the simulation box center with a mean displacement of $\overline{\Delta r\x}=$\SI{0.23}{\nano\meter}. The PNIPAM distribution for each monomer has been evaluated with respect to the COM of BIS. The closest monomers to the cross-linker distribute in a bimodal fashion, stemming from multiple possible side-chain--side-chain and cross-linker--side-chain interactions. This effect averages out for chain monomers further distant from the linker resulting in Gaussian distributions. The distances of two adjacent monomers is roughly \SI{0.25}{\nano\meter}. The distribution of all monomers together (i.e., averaged over a spherical shell $\propto \rho\mer(r)4\pi r^2 dr 
$ including all chains) shows a  plateau region between \SI{1.5} and \SI{2.5}{\nano\meter}, see \cref{fig:wat_distro}b. In this range we find an almost constant monomer density, which can be used to evaluate the intrinsic adsorption per chain monomer. We thus define  the number of monomers in an interval $[r_1,r_2]$ as
\begin{align}
 N\mer(r_1,r_2)=\int_{r_1}^{r_2}\rho\mer(r)4\pi r^2dr \label{eq:N_mer}
\end{align}
with $\rho\mer(r)$ the radial density of PNIPAM monomers with reference to the cross-linker. The number has the upper bound $N\mer(0,\infty)=48$, which is the total number of PNIPAM monomers in the system.

\subsection{Solute adsorption \label{sec:separate_ads} in high dilution}

The adsorption of solute inside the hydrogel of volume $V_{\mbox{\tiny gel}}$  (including the containing water) can be calculated from the radial distribution as
\begin{align}
  \G=& \rho_0 \int\displaylimits_{V_{\mbox{\tiny gel}}}\left[\gcos(r)-1\right]dV, \label{eq:adsorption}
\end{align}
which is a Kirkwood--Buff integral\cite{kirkwood1951statistical,ben2009molecular} counting the excess (or deficit) number of solutes with respect to bulk concentration $\rho_0$ and depends on the volume $V_{\mbox{\tiny ex}}$ excluded by the polymer. The adsorption $\G=0$ refers to the scenario at which the attractive solute--polymer interaction fully compensates for the steric exclusion $-\rho_0V_{\mbox{\tiny ex}}$.

Transferring this concept to our setup (\cref{fig:setup}), for which radial density profiles $g(r)$ of the solutes (\cref{fig:gcos}a, b) are measured from the COM of BIS, we define the partial adsorption $\G(r_1,r_2)$ counting excess solutes in the interval $\left[r_1,r_2\right]$, reading
\begin{align}
  \G(r_1,r_2)=&\rho_0\int_{r_1}^{r_2}\left[\gcos(r)-1\right]4\pi r^2 dr,  
\end{align}
and can scan the adsorption in different domains with respect to the cross-linker as shown in \cref{fig:wat_distro}. The total adsorption, i.e., $\G\tot=\G(0,\infty)$, is not only solute-specific but also depends on the number of monomers and the geometry. To separate the effects of our particular system setup, we distinguish now between three different contributions, stemming from the cross-linker ($\G\x$), linear chains ($\G_{\mbox{\tiny chain}}$), and chain terminal ends ($\G_{\mbox{\tiny end}}$). We will determine them by classifying different adsorption domains {\bf I, II}, and {\bf III},  and bulk phase ({\bf IV}),  as depicted in \cref{fig:wat_distro}. The total adsorption can be written as
\begin{align}
 \G\tot=\G\x+\G_{\mbox{\tiny chain}}+\G_{\mbox{\tiny end}}. \label{eq:gamma_tot}
\end{align}
 In our setup, the total adsorption is dominated by the chain contributions due to the numerous PNIPAM monomers compared to only one cross-linker. The contribution of the chain ends $\G_{\mbox{\tiny end}}$ is of lesser importance for this work. Dangling ends in hydrogels are very common but usually not of high concentration. It can be computed once the chain and cross-linker terms have been determined. {In our setup, however, the calculated values of $\G_{\mbox{\tiny end}}$ cannot be interpreted in a meaningful way due to the position restraints, which locally alter the relative water--polymer dynamics, in other words, disable the `dangling' behavior of such terminals. }

The equilibrium bulk concentration $\rho_0$ depends on the simulation box size and the binding affinity. It is convenient to define an infinite-dilution solute-specific adsorption coefficient that does not depend on concentration via
\begin{equation}
\G^* = \frac{\G}{\rho_0}.
\end{equation}
The adsorption coefficients $\G^*$ have the units of volume (nm$^3$), and correspond to highly diluted cases ($\rho_0\to0$), in which solute--solute interactions can be neglected. In the case of the highly water-soluble methanol, tested with 30 molecules, and simple ions (three pairs), solute--solute interactions play a minor role for the adsorption. Thus they are to some extent considered as very diluted scenarios and are comparable to single-solute simulation results. 
The different adsorption coefficients for all compounds are summarized in \cref{tab:sim_time}.

\subsubsection{Adsorption per chain monomer}

\begin{figure*}
  \includegraphics{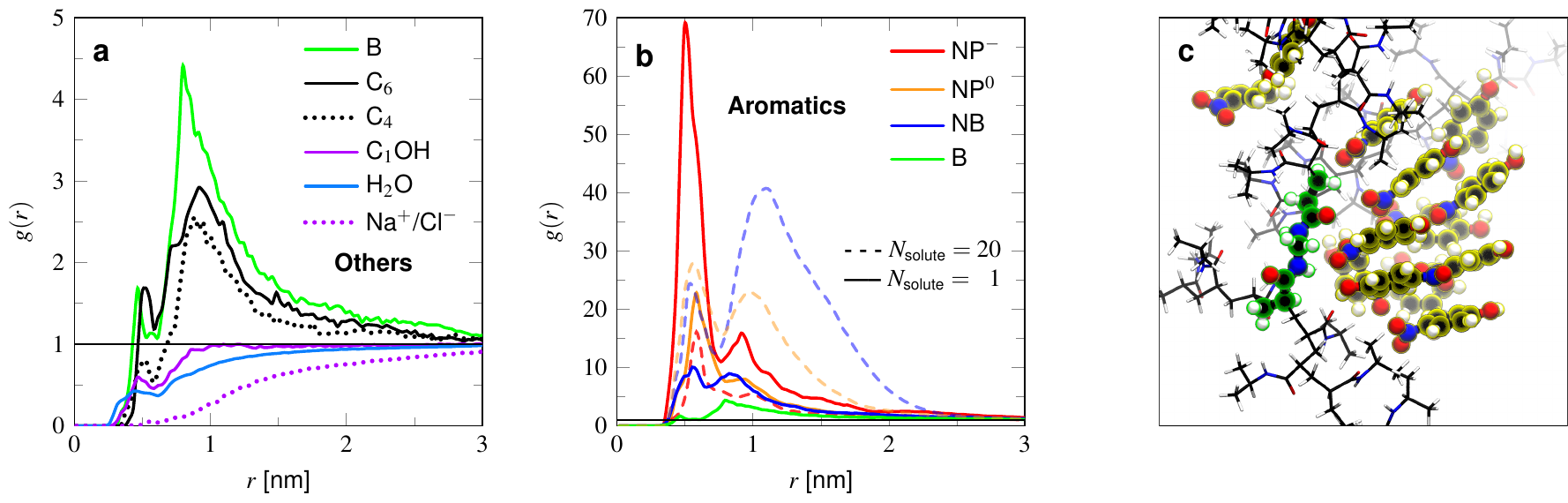} 
 \caption{({\bf\sffamily a}) Normalized radial density profiles of the solutes with respect to the COM of the BIS-linker for the low-adsorption species (C$_6$, C$_4$, C$_1$OH, Na$^+$, Cl$^-$),  \label{fig:gcos} plus benzene ({\color{green}\bf-----} B) and water ({\color{mycolorwat}\bf-----} H$_2$O) profiles for comparison. Note that in the case of methanol ({\color{mycolor2}\bf-----} C$_1$OH)  and sodium chloride ({\color{mycolor2}$\mathbf{\cdots\cdot}$} $\text{Na}^{+}/\text{Cl}^{-}$) simulations were carried out with 30 molecules or 3 ion pairs, respectively. Panel ({\bf\sffamily b}) shows much stronger adsorbing molecules BZ, NB, NP$^0$, NP$^-$, all of aromatic nature, in the high dilution limit ({\bf-----}) and more concentrated solutions with a total of 20 ({\bf- - -}) molecules (and 20 Na$^+$ counterions in the case of NP$^-$) per simulation. ({\bf\sffamily c}) Simulation snapshot showing a stacking of NP$^0$ in the cross-link region. BIS and NP$^0$ are 
highlighted green and yellow respectively, PNIPAM is shown in licorice representation. Note that the aggregation of the solutes usually looks  less ordered than presented due to thermal fluctuations.}
\end{figure*}

The chain contribution to the solute adsorption is what one would expect from a single isolated linear PNIPAM chain, i.e., in the absence of the cross-linker and any other chains close by. It can be described by the adsorption per monomer $\G\mer$ and with $N\mer(0,\infty)=48$ in our setup, this yields
\begin{align}
 \G_{\mbox{\tiny chain}}=N\mer\G\mer \label{eq:gamma_mer}
\end{align}
The adsorption per monomer is evaluated from the chain domain (where the BIS and end effects are negligible, see {\bf II} in \cref{fig:wat_distro}), i.e., $r\in[r_a,r_b]$, reading
\begin{align}
  \G\mer=\frac{\G(r_a,r_b)}{{N\mer}(r_a,r_b)}.
\end{align}

We now compare the adsorption of the solutes to the PNIPAM chain, listed in \cref{tab:sim_time} and visualized in \cref{fig:master_plot}. The results can further be compared with the density profiles (\cref{fig:gcos}{a, b}). We start with the examination of the rather weakly adsorbing species (\cref{fig:gcos}{a}). Methanol is the smallest probe molecule tested and is highly soluble in water and shows a rather low binding propensity. It is in fact slightly attracted to the polymer, but this cannot compensate the volume exclusion effect of PNIPAM and thus its adsorption coefficient is of negative value. {Methanol's preferential adsorption has already been reported in experiments\cite{winnik1992consolvency} and other simulations,\cite{dalgicdir2017computational,Mukherji2016,pang2010solvation,rodriguez2015mechanism} studying primarily the cononsolvency of PNIPAM in water--methanol mixtures.  }

Sodium and chloride have the lowest binding affinity to the hydrogel, which has already been shown in previous simulations of isolated chains. \cite{matej,du2010effects} As expected, simple well-hydrated ions are repelled from low dielectric (less polar) regions. 

The two probed alkanes, butane and hexane, have very similar profiles. The bigger hexane shows slightly higher adsorption than butane owing to the larger surface area, which facilitates hydrophobic interactions with apolar groups of the polymer chains. The very same argument does not hold when comparing with benzene. Benzene has about the same size as the alkanes, but shows higher binding affinity than the larger hexane. On the molecular level, the adsorption mechanism looks similar. Benzene and hexane tend to preferentially make contact with  hydrophobic parts of the polymer.

Comparing the aromatic compounds, which are roughly of equal size, we find the adsorption generally increases with the polarity of their substituents. The order according to their polarity, starting with the apolar benzene, is B$\to$NB$\to$NP$^0\to$NP$^-$, cf.\ \cref{fig:gcos}{b}.  All of them own a hydrophobic aromatic ring, interacting with the hydrogel described as in the benzene case. 
With one polar substituent for benzene, namely the nitrobenzene, the adsorption is more than doubled. This stems from additional hydrogen bonding {\cite{luzar1996hydrogen}} between the nitro-oxygens and the polymer's amide hydrogens. The very same interaction mechanism applies to NP$^0$ and NP$^-$. The extra hydroxy tail in the case of nitrophenol (NP$^0$) does not lead to a significant change of the adsorption to PNIPAM. On the one hand, the OH group can interact with the polymer's amide group and on the other hand, increases the water solubility. These two effects seem to compensate for NP$^0$ adsorption to the chains, such that the adsorption is similar to the one for NB. 

The deprotonated and hence charged NP$^-$ is the best adsorbing compound tested. The deprotonation leads to a redistribution of the electronic density, increasing the polarity of the whole molecule. The higher charging of the nitro-oxygens as well as the O$^-$ tail stabilize the contacts with PNIPAM's amide hydrogens, resulting in a roughly two times higher adsorption coefficient compared to NB and NP$^0$.

\subsubsection{Effects of the cross-linker}
 
\begin{figure}[!t]
  \centering
  \includegraphics{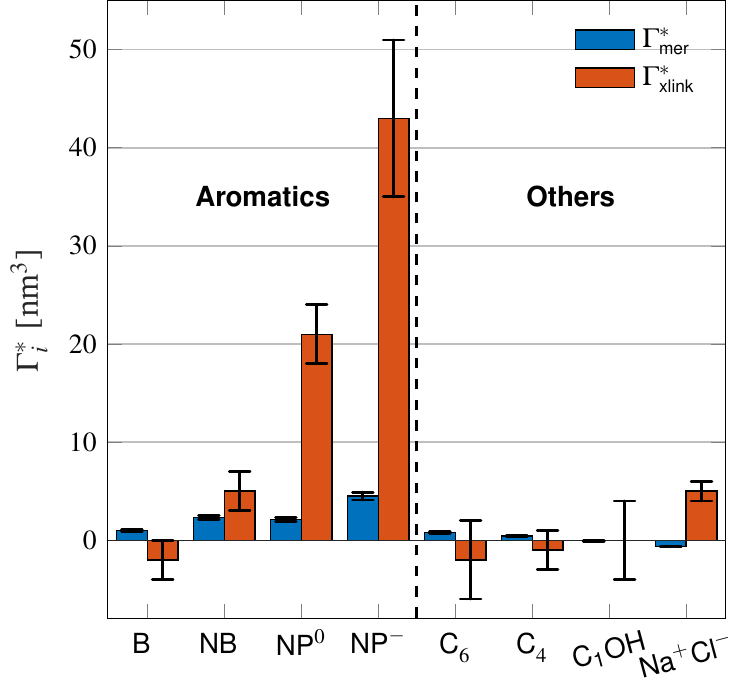}
    
 \caption{Adsorption coefficients ($\G^* = \G/\rho_0$) quantifying the binding affinities of different solutes to one PNIPAM monomer $\G^*\mer$ and the cross-linker $\G^*\x$  in the infinite dilution regime, as summarized in \cref{tab:sim_time}. The adsorption to the monomers increases with size, e.g., compare C$_4$ $\to$ C$_6$. The aromatic molecules show the highest adsorption, which is magnified by increasing polarity B $\to$ NB $\to$ NP$^0 \to$ NP$^-$, coinciding with a cross-linker-enhanced binding affinity.\label{fig:master_plot}}
\end{figure}

The contribution of the cross-linker in \cref{eq:gamma_tot} is obtained by integrating the solute radial density distribution from the COM of the BIS-linker up to the onset of the chain domain $r_a$ (cf.\ domain {\bf I} in \cref{fig:wat_distro}) and subtracting the estimated linear chain contribution therein, formally written as
 \begin{align}
    \G\x=\G(0,r_a)-\G\mer N\mer(0,r_a) .
 \end{align}
 This quantity combines specific interactions of the solute with the BIS-linker and the more complex many-body effects resulting from the higher concentration of PNIPAM monomers.
 
Note that in comparison to PNIPAM monomers, BIS has two amide groups and no isopropyl groups, thus creating a more hydrophilic environment than the chains. The apolar compounds, B, C$_4$, and C$_6$ show a slightly negative binding affinity in the cross-link region. In contrast, see again \cref{tab:sim_time} and \cref{fig:master_plot}, the solute adsorption $\G^*\x$ increases with polarity, where nitro-aromatic solutes are especially attracted.  Nitrobenzene shows a more than doubled adsorption to the cross-link region when compared to bare chain monomers. The NP$^0$ and NP$^-$ adsorption per cross-linker is even tenfold higher. The binding mechanism is similar to the single chain adsorption. The numerous amide hydrogen combinations make it very probable for the nitro-oxygens to find binding partners. 

\begin{figure}[!t]
\includegraphics[width=\linewidth]{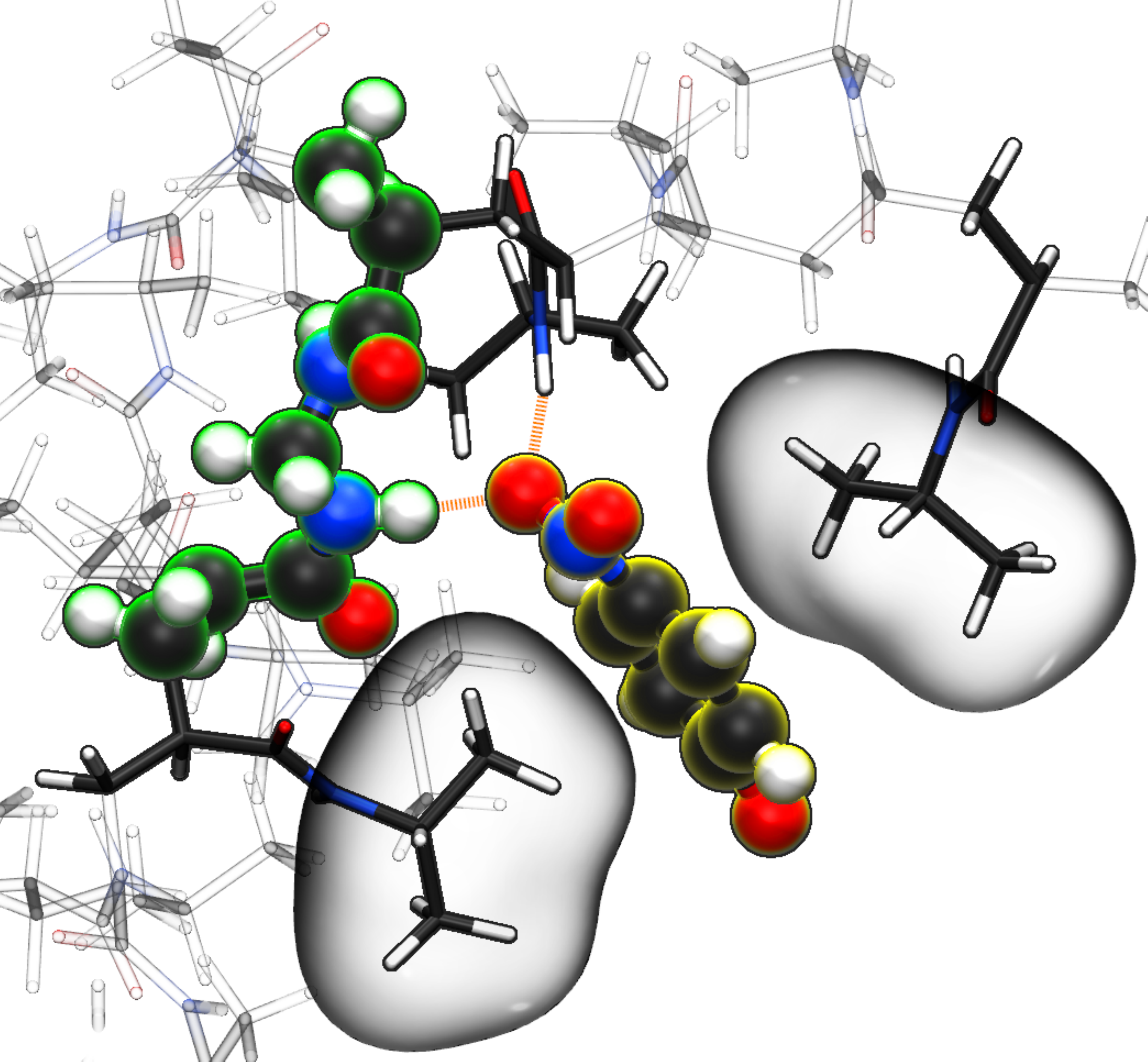}
 \caption{Simulation snapshot of NP$^-$ (illuminated yellow) benefiting from several possible interaction sites in the BIS-linker (illuminated green) proximity, serving as an illustrative explanation for the strong adsorption amplification due to the cross-linker (see \cref{tab:sim_time} and \cref{fig:master_plot}). Nitro-oxygens of NP$^-$ can form hydrogen bonds  (dashed orange lines {depict potential hydrogen bond formation in this configuration}) with numerous amide hydrogens, whereas the non-polar aromatic ring is surrounded by hydrophobic environment, i.e., isopropyl groups, (highlighted by bubbles) of the flexible PNIPAM side-chains. An aromatic ring--backbone contact has been observed but less frequently than the presented scenario. NP$^-$ can stay in such a conformation (with interchanging binding sites) for several tens of nanoseconds. \label{fig:truelove}}
\end{figure}

As already stated for the chain adsorption, NP$^-$ has the most polar nitro group resulting in the strongest adsorption coefficients in this study. Examining the simulation trajectories, we repeatedly found NP$^-$ in the location shown in \cref{fig:truelove}. One or both of the nitro-oxygens couple (forming hydrogen bonds) with two to three hydrogens from the amide groups: one from  BIS and one or two from the PNIPAM monomers. Additionally, the hydrophobic isopropyl groups or the backbone of PNIPAM can contact, almost embed, the aromatic ring, enhancing the stability of such an adsorbed state. The same mechanism has been observed for NB and NP$^0$, but the higher partial charges of NP$^-$ promote the binding.

\subsubsection{Finite concentration effects}

The adsorption in the low density case can differ from scenarios with higher concentrations owing to solute--solute interactions. This was tested with nitrobenzene, nitrophenol, and nitrophenolate, using twenty molecules per species, where we moved by an order of magnitude from the 2--3 mM concentration range up to 30--40 mM.  The concentrations and local adsorption results are also summarized in \cref{tab:sim_time}, while density profiles and structures are shown in \cref{fig:gcos}b, c, where we compare them to the low-density limit. We find for all tested solutes significant collective effects. The linear dependence of the adsorption on $\rho_0$ thus only holds for very low concentrations, in the millimolar regime. 

The least polar compound among them, nitrobenzene, shows the most substantial amplification of binding to the cross-linker at higher concentrations. NB is known to form NB--NB pairs and stacks of the aromatic rings,\cite{hunter2001aromatic} resulting thus in positive cooperativity for local adsorption (refer to earlier work\cite{matej} for further explanation). {Note that the bulk concentration of $\rho_0=\SI{28\pm2}{\milli\molar}$ might have exceeded the solubility of nitrobenzene in water. At \SI{298}{\kelvin} the experimental value is $\SI{16}{\milli\molar}$, but computer simulations can overestimate the solubility ($\SI{115}{\milli\molar}$).\cite{jorgensen2000prediction}}

NP$^0$ also performs stacking (\cref{fig:gcos}c), but due to its additional OH-tail, it has a higher water solubility and is thus less probable to aggregate. In stark contrast, the NP$^-$ adsorption to the whole network unit drastically drops at higher concentrations due to their electrostatic repulsion. This is an example of strong negative cooperativity of adsorption at higher concentrations. Note that a real hydrogel may change in size (in particular close to its {VPTT}) because of the solute--polymer interactions and that solutes may occupy a non-negligible volume, which in return limit the solute adsorption.\cite{hofmann2012dynamics,coughlan2006drug,kim2017cosolute,du2010effects}

\subsection{Partitioning in a hydrogel \label{sec:partition}}

The adsorption coefficients retrieved in this work can be used to estimate the resulting solute partition ratios in swollen PNIPAM--BIS hydrogels, { which will be compared with experimental data in this section}. To this end, we extrapolate our results using an idealized, mean-field model of a large network.
 \subsubsection{{Idealized hydrogels}}
As a start, the solute partition ratio is determined by the outside bulk concentration $\rho_0$ and the concentration inside the hydrogel $\rho_{\mbox{\tiny in}}$, namely 
\begin{align}
 K=\frac{\rho_{\mbox{\tiny in}}}{\rho_0},\qquad \mbox{with } \quad  \rho_{\mbox{\tiny in}}=\frac{N_{\mbox{\tiny in}}}{V_{\mbox{\tiny gel}}}. \label{eq:partitioning0}
\end{align} 
Here, $V_{\mbox{\tiny gel}}$ is the volume of the entire hydrogel including the water and should not be confused with the excluded volume $V_{\mbox{\tiny ex}}$.

The number of particles $N_{\mbox{\tiny in}}$ inside the gel can be assessed using the total solute adsorption as
\begin{align} 
 N_{\mbox{\tiny in}}=\G_{\tot}+\rho_0V_{\mbox{\tiny gel}}. \label{eq:N_in}
\end{align} 
If the hydrogel has no net effect, i.e., $\G_{\tot}=0$, the concentration inside the gel is equal to the bulk value $\rho_0$, and the particles inside the gel account to $\rho_0V_{\mbox{\tiny gel}}$. Neglecting the effects of dangling (terminal) ends, the adsorption can be assumed to be the sum of single chain adsorption and the effect of all cross-linkers, yielding
\begin{align}  
 \G\tot=N\mer\G\mer+N\x\G\x, \label{eq:adsorption_0}
\end{align}
where $N\mer$ and $N\x$ stand for the number of PNIPAM monomers and BIS-linkers, respectively.
Plugging all ingredients into \cref{eq:partitioning0}, the solute partition ratio can be expressed as
\begin{align}
 K=1+{\rho\mer}\left( \G^*\mer+\alpha\G^*\x\right), \label{eq:partitioning}
\end{align}
with the PNIPAM monomer concentration $\rho\mer={N\mer}/{V_{\mbox{\tiny gel}}}$, the BIS-to-PNIPAM monomer ratio $\alpha={N\x}/{N\mer}$, and adsorption coefficients  $\G^*={\G}/{\rho_0}$. 

\begin{figure}[t]

\includegraphics{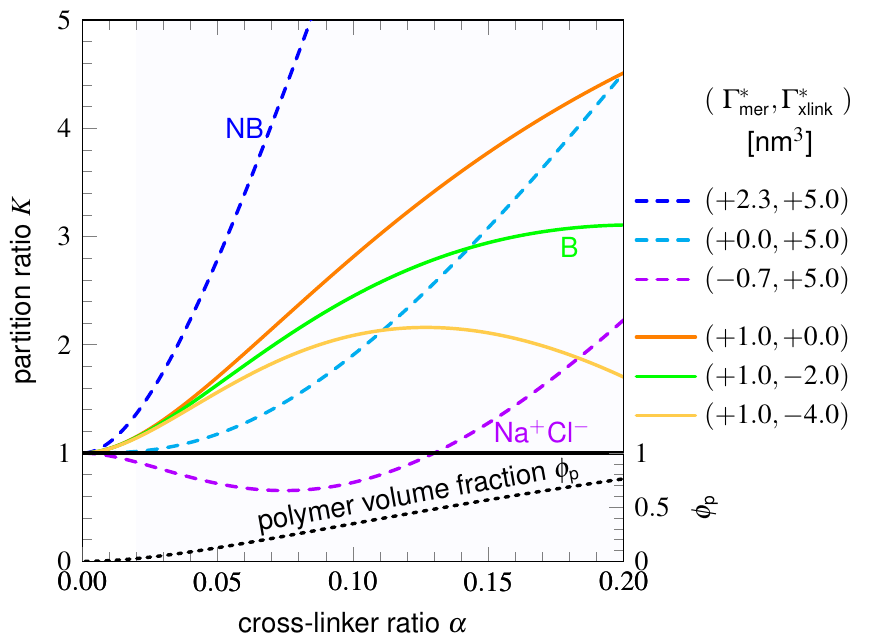}

 \caption{Partition ratios for different solutes in an ideal diamond-lattice polymer network (see \cref{sec:appB}) as a function of cross-linker ratio. The curves result from a competition between the adsorption coefficients $\G^*\mer$ and $\G^*\x$. Dashed lines have the same cross-linker adsorption, solid lines have identical monomer adsorption. Nitrobenzene (NB) has an overall positive binding affinity and $K$ is strictly an increasing function of $\alpha$. Benzene (B) has a positive adsorption to the chain monomers but a slightly negative cross-linker effect, resulting in a maximum value at $\alpha=0.2$. The Na$^+$/Cl$^-$ pair is the opposite case, it has a negative chain adsorption coefficient but a positive one for the linker, exhibiting a partitioning minimum. The orange line presents the same case but with zero cross-linker effect ($\G^*\x=0$) and it has a linear relation to the volume fraction.  The black dashed line presents the polymer volume fraction $\phi_p$ for this idealized diamond network 
and its scale is on the right.\label{fig:partitioning}}
\end{figure}

Considering now a defect-free diamond lattice network architecture\cite{perez2018maximizing,schneider2002swelling} of the hydrogel, we can deduce the functional form of the monomer concentration vs.\ the cross-linker ratio, i.e., $\rho\mer\to \rho\mer(\alpha)$, see \Cref{sec:appB}.
For different adsorption coefficient pairs ($\G\x^*$, $\G\mer^*$) and in dependence on the cross-linker ratio, $K$ is visualized in \cref{fig:partitioning}. We find that $K$ and $\alpha$ can have a non-linear relation and even a non-monotonic behavior.  The reason is that higher cross-linker ratios directly enhance the influence of $\G\x^*$, and additionally, as already discussed, increase the PNIPAM concentration $\rho\mer(\alpha)$, promoting the influence of $\G\mer^*$.  If now $\G\x^*$ and $\G\mer^*$ have even different signs, i.e., a solute, for example, is preferentially desorbed from the polymer but adsorbed by cross-linker then naturally non-monotonic behavior must occur. 

Typical values for cross-linker ratios in experiments range from roughly $0.02$ to $0.2$. In our model (see \cref{sec:appB}), this corresponds to volume fractions ranging from approximately $0.03$ to $0.75$ with an almost linear relation to $\alpha$ in this interval. Thus the plotted region in \cref{fig:partitioning} is quite reasonable for demonstrating the non-linear and non-monotonous $\alpha$-dependencies of the partition ratio. In particular, with positive adsorption coefficients, like all nitro-aromatics have, we find that the partition ratio monotonically increases with larger cross-linker ratios, exemplified by nitrobenzene in \cref{fig:partitioning}.

{Selective solute--cross-linker binding affinities are not solely responsible for an increasing partition ratio when increasing the cross-link ratio. As an illustration, we show scenarios of hypothetical solutes that have either zero adsorption to cross-linkers or zero adsorption to the chains.} For small values of $\alpha$, the coefficient $\G\mer^*$ has greater impact on partitioning increase than $\G\x^*$. In the case of benzene (similarly hexane and butane), where we find a positive chain adsorption, but repulsion from the cross-linker, the partitioning reaches a plateau at $\alpha=0.2$. Weaker chain adsorption, or stronger cross-linker repulsion can lead to a maximum in the plotted range, which is exemplified by the hypothetical solute with $\G\mer^*=\SI{1}{\nano\meter^3}$ and $\G\mer^*=\SI{-4}{\nano\meter^3}$. The very opposite case, i.e., cross-linker affinity in combination with chain avoidance, as we find for the tested ion pair Na$^+$/Cl$^-$, exhibits a  minimum. 

Summing up, the adsorption coefficients $\G\mer^*$ and $\G\x^*$ determine the gradient and concavity of the solute partitioning in dependence on the cross-linker ratio, assuming homogeneous and diamond lattice-like network structure. We conclude that partitioning vs.\ the cross-linker ratio may be complex and non-monotonous, exhibiting minima and maxima and intercepting the $K=1$ line.

 \subsubsection{{Relating to experiments}}
 In real hydrogels, one has to be aware of additional effects. One strong assumption in our model is a rigid and homogeneous network structure, which in general is not the case in the real world. Though techniques have emerged to control the cross-linker density throughout the gel, \cite{acciaro2011preparation,still2013synthesis,meyer2005influence} the hydrogel structure is still subject to the randomness of the polymerization process and thus retains inhomogeneities. This may lead to nano/micro cavities within the gel and more complex network architectures than assumed, influencing polymer volume fraction and partitioning. 
 
Our investigation focuses on very low solute concentrations, though the response of the polymer to the penetrants might not be negligible. From experiments\cite{hofmann2012dynamics,coughlan2006drug,Kawasaki1996} it is known, that solutes may change the {hydrogel's VPTT}, which has additionally been demonstrated in computer simulations. \cite{kim2017cosolute,du2010effects}

{Nevertheless, our idealized approach tackling the partitioning does allow for an indirect comparison to experimental data. One experiment on the rate of the nitrobenzene reduction in an (N-isopropylacrylamide-co-acrylic-acid (PNIPAM-co-AAc)) nanoreactor\cite{farooqi2015effect}  shows an increase in the reduction rate with increasing cross-link (BIS) density, which is attributed to the higher nitrobenzene concentration inside the hydrogel. Parasuraman et al.\cite{parasuraman2012poly} used a very similar hydrogel (PNIPAM-co-AAc-BIS) and proved the increased dye uptake (Orange II) with increasing cross-link ratio. Both studies qualitatively support our findings. Theym, however, have in common that the initial increase of the reduction rate and the dye uptake, respectively, apparently saturates for higher cross-link degrees. This effect is not captured by our model and might result from steric hindrances, i.e., undersized pore/mesh size of the polymer architecture and already occupied adsorption sites for higher solute concentrations.}

{After qualitatively confirming the impact of the cross-link ratio on the partitioning of the aromatic compounds, we will now assess the comparison in terms of absolute values. Experimentally, partition ratios have been reported for several molecules containing aromatic rings. A study by Molina et al.\cite{molina2012study} retrieved $K$ in PNIPAM-BIS-hydrogels ($\alpha=2\%$) for probe drugs (tryptophan, propranolol chloride, dansyl chloride, methyl orange, riboflavin and ruthenium-tris(2,2'-bibyridiyl) dichloride), which contain two to six aromatic rings as well as polar and/or charged residues. The partition ratio, depending on the compound, ranges from roughly 4.6 to 10.}

{Comparing to our perfect network model, the much smaller NB and NP$^-$ show partition ratios  of approximately 1.3 and 1.8 respectively at $\alpha=2\%$, and 2.7 and 5.7 at $\alpha=5\%$ in the low dilution limit ($N_{\mbox{\tiny solute}}=1$). For $N_{\mbox{\tiny solute}}=20$, $K$ is about 3.7 at $\alpha=2\%$ for NB. It is expected, that larger molecules at higher concentrations, as established in the mentioned experiments, will lead to higher partition ratios\cite{kanduc2018transfer} and can thus be regarded as supportive of our results. Furthermore, the adsorption increase due to positive cooperativity (e.g., NB) has been shown for methylene blue in a superabsorbent hydrogel.\cite{paulino2006removal}}

{The salt partition ratios in 1\% cross-linked PNIPAM-BIS gels at room temperature have been reported\cite{Kawasaki2000} and amount to $K_{\mbox{\tiny LiCl}}=0.97\pm0.05$, $K_{\mbox{\tiny KCl}}=0.91\pm0.05$ and $K_{\mbox{\tiny NaCl}}\approx 0.95$, i.e., just below unity as in our prediction for sodium chloride.} 

{However, the non-monotonicity of the partition ratio in dependence on the cross-link ratio predicted by our model has not been reported by experimentalists so far and is yet to be tested. }
\section{Concluding remarks}

We investigated the effects of cross-linking on solute adsorption in swollen PNIPAM hydrogels by means of explicit-water MD simulations at $T=\SI{290}{\kelvin}$, i.e., below the PNIPAM collapse transition temperature. We considered a generic hydrogel subunit consisting of one BIS-linker and four PNIPAM chains, which was kept in tetrahedral geometry by position-restrained backbone terminals. By subdividing the radial distance from the central cross-linker, we classified four different adsorption regions according to the polymer's prevalent features, namely the cross-linker region, the linear chain region, the chain terminal region, and the bulk solvent domain. We evaluated the adsorption of different solutes, representing typical charged, polar and nonpolar molecular compounds. 

Comparing the cross-linker and monomer effects on adsorption, we find different scenarios. Apolar species show small attraction to chain monomers and slight repulsion from the cross-linker region. Sodium chloride behaves the opposite, it has negative chain adsorption but is attracted towards the cross-linker. The strongest adsorbing solutes, the nitro-aromatics, adsorb to all parts of the polymer and show the highest binding affinity, which is promoted by hydrophobic interactions between the aromatic ring and PNIPAM's isopropyl groups as well as by hydrogen bonds between the nitro-oxygens and the amide groups. The adsorption at the cross-linker relative to a single PNIPAM monomer adsorption, spans from $ \G\x/\G\mer\approx2$ (NB) to $ \G\x/\G\mer\approx10$ (NP$^0$, NP$^-$). This indicates that the cross-linker can significantly enhance the overall adsorption to the network unit. Hence, for solutes that have a significant affinity to PNIPAM chains already, the dense cross-linker region, where many-body attractions 
are at play,  amplifies the local adsorption by even an order of magnitude. Thereby we confirm the `vertex trapping' effect that has been first reported in generic coarse-grained simulations of polymer networks.\cite{Zhang2015,hansing2016nanoparticle,hansing2018hydrodynamic,hansing2018particle,hansing2018particle2} 

In the case of the nitro-aromatics, we furthermore performed simulations with higher solute concentrations to estimate cooperative adsorption effects. Nitrobenzene shows enhanced aromatic stacking at the cross-linker and the adsorption is elevated in a superlinear fashion with increasing concentration. Nitrophenol shows similar, positive cooperativity but a less pronounced behavior. For both NB and NP$^0$ the cross-linker promotes higher positive cooperation effects than single chains. In contrast, in the case of the negatively charged nitrophenolate, we observe less adsorption at higher concentrations due to negative cooperativity stemming from the electrostatic repulsion.

The adsorption coefficients  for cross-linker and chain monomers in the low concentration regime were used to estimate partition ratios of the solutes within an idealized, homogeneous diamond-lattice macrogel, {which allowed a comparison with experimental findings. In our model,} we found that highly adsorbing substances like nitro-aromatics have a partition ratio ranging from $2$ to $5$ at a cross-linker concentration of 5$\%$.  Solutes with adsorption coefficients of opposite signs may show non-monotonic behavior as a function of the cross-linker ratio: Positive/negative chain adsorption and negative/positive cross-linker adsorption leads to a maximum/minimum in the partition ratio.  These yet poorly known features should be considered in future experiments and modeling of hydrogels as they play an important role for the fine-tuning of solute uptake within the needs of the desired function and application.

\section*{Conflicts of interest}
There are no conflicts of interest to declare.

\section*{Acknowledgement}
The authors would like to thank Won Kyu `Q' Kim  and the whole ERC team for fruitful discussions. This project has received funding from the European Research Council (ERC) under the European Union's Horizon 2020 research and innovation programme (grant agreement no.\ 646659). M.K. acknowledges the financial support from the Slovenian Research Agency (research core funding no.\ P1-0055). The simulations were performed with resources provided by the North-German Supercomputing Alliance (HLRN). The authors acknowledge support by the state of Baden-W\"urttemberg through bwHPC.

\begin{appendix}
 \section{Force field parameters}
 \label{sec:appA}
 From our ab-initio calculations with the Gaussian 09 software,\cite{g09} we obtain partial charges for the cross-linker and monomers, which are very similar and in good agreement with the OPLS-QM2 force field.\cite{janOPLSQM2} Due to the chemical similarity of BIS and PNIPAM, we apply the PNIPAM's partial charges of the OPLS-QM2 force field for BIS atoms as well. The partial charges from the original OPLS-AA\cite{jorgensen_opls_1988} and OPLS-QM2 force fields as well as our results are shown in \cref{tab:charges}. The standard OPLS force field is used for all remaining parameters like potentials, masses, etc. The hitherto undefined  N--C--N angle and C--N--C--N dihedral potentials for the cross-linker were adopted from the OPLS C--C--C angle and C--N--C--C dihedral parameters, respectively.
\begin{table}[h]
\footnotesize
\sffamily
\centering
\caption{Partial charges (in unit charges) of different force fields and results from own ab-initio (Hartree--Fock) calculations on the 6-31G(d)-level with electrostatic potential fitting for BIS and PNIPAM. Chemical structure with the different atom classification are depicted in \cref{fig:charges}. BIS charges of our calculations were retrieved by analyzing 158 conformations of a molecule consisting of one BIS cross-linker and four PNIPAM monomers attached. PNIPAM charges were obtained from 36 different single chains consisting of  11 monomers. In this work we use the OPLS-QM2 partial charges for PNIPAM and BIS. The charges for C$^6$ and H$^6$ atoms in BIS were chosen similar to the C$^4$ and H$^4$ charges in PNIPAM and with overall electroneutrality of the cross-linker in mind.\label{tab:partchargeQM} {A small test simulation with slightly redistributed partial charges of the central C$^6$H$^6_2$-group (C$^6=0.1$,H$^6=0.15$,C$^3=0.56$) did not change the nitrobenzene adsorption affinity. The overall positive partial charges of these atoms (C$^6$H$^6_2$) and the many-body interactions (solute--linker--monomers) dominate} }

 \begin{tabular}{l S S S S S}
 & {\footnotesize OPLS-AA\cite{jorgensen_opls_1988}} &\multicolumn{2}{c}{\footnotesize HF-6-31G(d)} & {\footnotesize OPLS-QM2\cite{janOPLSQM2}} & {\footnotesize charges used}  \\
  & & \multicolumn{1}{r}{\footnotesize BIS} & \multicolumn{1}{r}{\footnotesize PNIPAM} & {\footnotesize PNIPAM} &{\footnotesize in this work} \\
\hline 
\rule{0ex}{2ex}C$^1$  & -0.12&  -0.22 & -0.20 & -0.18& -0.18\\
H$^1$ & 0.06& 0.07 & 0.08 & 0.09 & 0.09 \\
C$^2$ & -0.06&  -0.13 & -0.07 & 0.00  & 0.00  \\
H$^2$ & 0.06 & 0.06 & 0.05 & 0.05 & 0.05 \\
C$^3$ &  0.50& 0.81  & 0.80 & 0.50   & 0.50   \\
O & -0.50 & -0.61 & -0.65 & -0.57 & -0.57 \\
N & -0.50& -0.69 & -0.84 & -0.57 & -0.57 \\
H$^{\text{\tiny N}}$ & 0.30& 0.39 & 0.39 & 0.33  & 0.33  \\
C$^4$ & 0.14& {-}  & 0.64 & 0.36 & 0.36 \\
H$^4$& 0.06 & {-}& -0.01 & 0.06 & 0.06 \\
C$^5$& -0.18 & {-} & -0.52 & -0.32 & -0.32 \\
H$^5$& 0.06 & {-} & 0.12 & 0.08 & 0.08 \\
C$^6$& {-} & 0.06 &{-} & {-} & 0.40 \\
H$^6$& 0.06 & 0.14 & {-} &{-} & 0.06 \\

\end{tabular}\label{tab:charges}

\end{table}
\begin{figure}[h]
\includegraphics[width=\linewidth]{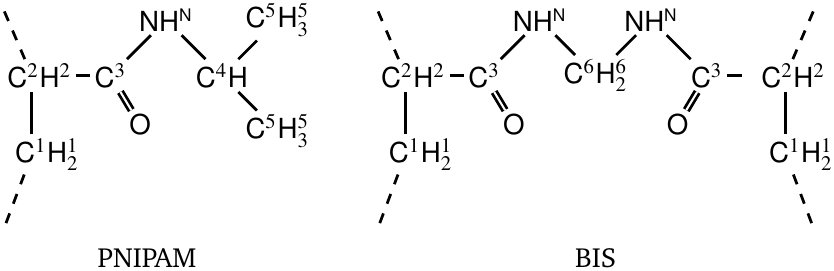}

\caption{Chemical structures of PNIPAM and BIS. The superscript indices differentiate atoms regarding their partial charges, shown in \cref{tab:charges}. \label{fig:charges}}
\end{figure}

 Recapturing the results for the adsorptions around single chains reported by our group before,\cite{matej} the reader may compare the standard OPLS-AA force field for PNIPAM polymers with the further optimized OPLS-QM2 version\cite{janOPLSQM2} employed in this work. 
 The qualitative trends, i.e., size and polarity dependence as well as the strong nitro aromatic binding affinity remain. However, comparing absolute numbers for aromatic compounds, the standard force field\cite{matej} shows a roughly twice as high adsorption coefficient to the chain monomers. We attribute this effect to more polar amide groups in the OPLS-QM2 force field and hence to more hydrophilic behavior.  
 %Local electrostatic interactions seem to play a greater role.
 Moreover, the adsorption of NP$^-$ to PNIPAM (in the OPLS-QM2 version) is significantly larger owing to its polar nitro group.

\section{Defect-free PNIPAM-BIS hydrogels}
\label{sec:appB}
We assume a homogeneous network with cross-linkers arranged in a perfect diamond lattice and thus connected with chains of an equal number of monomers. Given this perfect lattice, we can simply deduce from the geometry the monomer concentration as a function of the cross-linker ratio, $\rho\mer(\alpha)=n\mer(\alpha)/v_{\mbox{\tiny gel}}(\alpha)$. We can also assess the network's unit cell volume $v_{\mbox{\tiny gel}}(\alpha)$ and its PNIPAM monomer content $n\mer(\alpha)$. Such a unit cell contains eight cross-linkers and 16 chains, hence $n\mer(\alpha)=8/\alpha$, and the number of monomers between two associated cross-linkers is $n_{\mbox{\tiny chain}}(\alpha)=1/(2\alpha)$. The distance between two cross-linkers is then given by
\begin{align} 
 \ell_{\mbox{\tiny xx}}(\alpha)= {\ell_{\mbox{\tiny xlink}}}+n_{\mbox{\tiny chain}}(\alpha)\lambda\Delta L_{\mbox{\tiny c}}\mbox{,}
\end{align}
where $\ell_{\mbox{\tiny xlink}}=\SI{0.22}{\nano\meter}$, the effective length contribution of one BIS-linker, which is an average value retrieved from simulations solving $L={\ell_{\mbox{\tiny xlink}}}/2+{\ell_{\mbox{\tiny ee}}}$. The unit cell volume reads 
\begin{align}
v_{\mbox{\tiny gel}}(\alpha)=8^{\frac{3}{2}}\sin^3\left(\frac{\theta}{2}\right)\ell_{\mbox{\tiny xx}}^3(\alpha)
\end{align}
with $\theta\approx109.5\degree$ being the angle between any pair of adjacent chains. Eventually, the monomer concentration as well as the solute partition ratio, \cref{eq:partitioning}, are fully defined.

By knowing $n\mer(\alpha)$, we can further approximate the polymer volume fraction, which allows us to identify physically meaningful values of $\alpha$ and $K$. Using the water profile (\cref{fig:wat_distro}) we extract the excluded volume per monomer as $V_{\mbox{\tiny ex,mer}}=\SI{0.167}{\nano\meter^3}$ and assume the BIS' volume to scale with the number of heavy atoms (without hydrogens) compared to a PNIPAM monomer (11.\ vs.\ 8), yielding $V_{\mbox{\tiny ex,xlink}}=\SI{0.230} {\nano\meter^3}$. The excluded volume in one unit cell is then $v_{\mbox{\tiny ex}}=8V_{\mbox{\tiny ex,xlink}}+n\mer(\alpha)V_{\mbox{\tiny ex,mer}}$ and the polymer volume fraction for different values of $\alpha$ is obtained as $\phi_{\mbox{\tiny p}}=v_{\mbox{\tiny ex}}/v_{\mbox{\tiny gel}}$~(\cref{fig:partitioning}).

\end{appendix}

\printbibliography

\end{document}